\documentclass[12pt,preprint]{aastex}

\usepackage{bm}

\shorttitle{TF Study}
\shortauthors{Hodge}

\begin{document}

\title{Relationship Between Cepheid and Tully-Fisher Distance Calculations}

\author{John C. Hodge\altaffilmark{1}}
\affil{Pisgah Astronomical Research Institute, 1 PARI Drive, Rosman, NC, 28772}
\email{scjh@citcom.net}

\author{Michael W. Castelaz}
\affil{Pisgah Astronomical Research Institute, 1 PARI Drive, Rosman, NC, 28772}
\email{mcastelaz@pari.edu}

\altaffiltext{1}{Visiting from XZD Corp., 3 Fairway St., Brevard, NC, 28712 }

\begin{abstract}
A correlation between (1) the difference between the Tully-Fisher calculated distance and Cepheid calculated distance for a target galaxy and (2) the magnitude and distance of galaxies close to the target galaxy is described.  The result is based on a sample of 31 galaxies that have published Cepheid distances with a wide range of characteristics and distances from 2.02 Mpc to 49.7 Mpc.  The energy impinging on a target galaxy from neighboring galaxies is related to the residual between Tully-Fisher and Cepheid distance calculations of the target galaxy.  These  relations have four different zones dependent on the value of the impinging energy.  The correlation coefficients are 0.77, 0.87, 0.81, and 0.98, respectively.  This relationship is of interest not only for its ability to reduce the Tully-Fisher to Cepheid distance residual but also because it suggests neighboring galaxies have a strong influence on the apparent luminosity of galaxies and it suggests the Tully-Fisher relationship assumption that the intrinsic mass to intrinsic luminosity ratio is constant among galaxies is valid for $z < 0.006$.
\end{abstract}
\maketitle
\section{INTRODUCTION}

The residual between the Cepheid distance calculation $D_c$ for a galaxy and the Tully-Fisher relationship (TF) distance calculation $D_{tf}$ for the galaxy varies among galaxies \citep{shan, shan2}.  The residual can be small for some galaxies or greater than the sum of the errors of each calculation for other galaxies.  The $D_{tf}$ to a target galaxy is determined by relating the 21-cm line width at 20 percent of the peak value $W_{20}$ corrected for the inclination between our line of sight and the target galaxy's polar axis $ i $ with the absolute magnitude $M$ in various wavelength bands \citep{tull}.  The $D_c$ is based on the observation that pulsating Cepheid stars in a galaxy have a $M$ related to their pulse period.  Cepheids obey very well defined relations and so make excellent standard candles \citep{binn}(page 415).  This paper uses $D_c$ as standard distances.  In this Paper, the influence of neighboring galaxies on $M$ is investigated for the first time.  A remarkably tight correlation is found relating the illumination of neighboring galaxies to ($D_c-D_{tf}$) for redshift $z < 0.006$.  Since this result was obtained with the assumption that the mass of a galaxy is constant over time, there may be a fundamental physical mechanism regulating the mass of galaxies.  Also, the TF assumption that the intrinsic mass to intrinsic luminosity ratio is constant among galaxies \citep{aaro} is valid for $z < 0.006$.

The object of this paper is to examine the effect of neighboring galaxies on $D_{tf}$.  Thereby, the residual between $D_c$ and $D_{tf}$ distances may be reduced.  The approach is to (A) develop an equation from the Principle of the Conservation of Energy describing the influence of neighboring galaxies on $D_{tf}$, (B) choose galaxies with published Cepheid distances, and (C) apply the model to the target galaxies.

\section{MODEL}

The light emitted from a galaxy is energy from the galaxy.  The Principle of the Conservation of Energy in a galaxy requires,
\begin{equation}
\Delta \sf {M} \it c^2+ E_{out} = E_{in} + \epsilon
\label{eq:1},
\end{equation}
where $\Delta \sf M$ is the rate of change of mass $\sf M$ in the target galaxy, $c$ is the speed of light, $E_{out}$ is the rate energy radiates from the target galaxy, $E_{in}$ is the rate of transmittal of energy from another galaxy into the target galaxy, and $\epsilon$ is the rate of intrinsic energy production from mass conversion processes in the target galaxy.

\citet{mclu} found that the mass in the central region of a galaxy is 0.0012 of the mass of the bulge.  \citet{fer5} reported that about 0.1\% of a galaxy's luminous mass is at the center of galaxies and that the density of supermassive black holes (SMB) in the universe agrees with the density inferred from observation of quasars.  \citet{merr, fer2} found similar results in their study of the relationship between the mass $\sf M_\bullet$ of a SMB and dispersion velocity $\sigma$ of a galaxy.  \citet{fer4} found a tight relation between rotation velocity $v_c$ and bulge velocity dispersion $ \sigma_c$ which is strongly supporting a relationship of the center force with total gravitational mass of a galaxy.  Either the dynamics of many galaxies are producing the same increase of mass in the center at the same rate or a mechanism exists to evaporate (convert to radiated energy) the mass increase that changes as the rate of inflow changes.  The rotation curves imply the dynamics of galaxies differ such that the former is unlikely.  \citet{merr2} suggested a feedback mechanism must exist in such a way that $\Delta \log (\sf {M_\bullet}$$ ) \approx 4.5 \Delta \log (\sigma )$.  Therefore, $\sf M$ is assumed to remain constant, the matter in an incremental cylinder shell of a galaxy is constant, and $\Delta \sf M = 0$.  Whatever the source of matter in a galaxy, the energy content of the inflowing matter equals $\epsilon$.  Constant mass in a galaxy and the observation of $\epsilon$ implies there is inflowing matter.

Another source of energy for a galaxy is from other galaxies as $E_{in}$.  The energy transfer from other galaxies can be (A) in the form of photons $E_a$, (B) in the form of potential energy $E_p$ from gravitational forces of other galaxies.  The $E_a$ component is dependant on the surface area cross section of the target galaxy presented to the other galaxy, the distance between galaxies, and the $M$ of the other galaxies.    The $E_p$ is a conservative field.  If the matter in an incremental cylinder shell of the target galaxy is constant, then $E_p = 0$.  The heat change caused by the movement of a galaxy as a whole in a potential field is assumed to be negligible in the B band.

Therefore, conservation of energy implies the only radiated energy observed is from re-radiated energy from neighboring galaxies and from $\epsilon$,
\begin{equation}
E_{out} \propto 10^{\frac{-M}{2.5}} = K_a E_a + K_e \epsilon
\label{eq:2},
\end{equation}
where $K_a$ and $K_\epsilon$ are proportionality constants of $E_a$ and $\epsilon$, respectively [herein called ``Conversion Efficiencies'' (CE)] and $E_{in} = K_a E_a$.  The CE's represent the fraction of energy transmitted in the measurement band of $M$ relative to the available energy.  Hence, $ K_a E_a$ is the observed luminosity originally from neighbor galaxies re-emitted from a target galaxy and $ K_e \epsilon $ is the observed intrinsic luminosity of the target galaxy.  Since $\epsilon$ will vary among galaxies, $ K_e \epsilon $ will also vary among galaxies.  The $K_a$ is a function of the target galaxy's absorption per unit surface area and the re-emitted light per unit absorption.  Since the HI mass to total mass ratio increases systematically from giant galaxies towards dwarf galaxies \citep{kara}, $K_a$ may change accordingly.  Therefore, it is reasonable to assume that the $K_a$ also changes with $\epsilon$.

Consider the ($D_c - D_{tf}$) as an error in $D_{tf}$.  The error rate is defined as the correction factor $C_f$,
\begin{equation}
C_f \equiv \frac{ D_c - D_{tf} }{ D_{tf} }
\label{eq:6},
\end{equation}
or, 
\begin{equation}
D_c = D_{tf} ( 1 + C_f)
\label{eq:7}.
\end{equation}

Consider the attenuation of energy is a constant factor $F_a$ of the light from the target galaxy.  The term on the left side of equation (\ref{eq:2}) is the luminosity calculated by the TF relation.  If $D_c$ is the true distance, the last term on the right side of equation (\ref{eq:2}) is the intrinsic luminosity that would be calculated using $D_c$.  Therefore, $D_{tf} = F_a 10^{-M/2.5}$ and $D_c = F_a K_e \epsilon$.  Substituting these relations into equation~(\ref{eq:6})and solving equations~(\ref{eq:2}) and (\ref{eq:6}) for $C_f$ yields,
\begin{equation}
C_f = \frac{K_a E_a}{K_e \epsilon - K_a E_a}
\label{eq:8}.
\end{equation}

A binomial expansion of the denominator of equation~(\ref{eq:8}) yields,
\begin{equation}
C_f = K_s E_a + K_i
\label{eq:9},
\end{equation}
where $K_i$ is the sum of second and  higher order terms and is approximately a constant and, 
\begin{equation}
K_s = \frac{K_a}{K_e \epsilon}
\label{eq:9a}.
\end{equation}

If $K_a \propto K_e \epsilon$ among galaxies, then a plot of $C_f$ versus $E_a$ will yield a linear relationship.  That is, for each $E_a$ there corresponds one and only one value of $C_f$.  Therefore, $C_f$ is a functional relationship of $E_a$.  The function may have discontinuities and gaps but it is still one relationship.

Galaxies with a measured $D_c$ can be used as calibration galaxies to calculate $K_s$ and $K_i$.  Then equations (\ref{eq:7}) and (\ref{eq:9}) can be used to calculate a distance corresponding to a $D_c$.

\section{DATA AND ANALYSIS}

The criteria for choosing galaxies for the analysis are (1) the galaxy has a published Cepheid distance, (2) the distance to the galaxy must be large enough that the Milky Way's contribution to $E_a$ is negligible, (3) the distance to the galaxy must be large enough that the contribution of the peculiar velocity to the redshift $z$ measurement is relatively small, and (4) either a published total apparent magnitude in the in the B band $m_b$ or a published total apparent corrected I-band magnitude $m_i$ must be available for both the target and close galaxies.

The data and references for the 31 target galaxies is presented in Table~\ref{tab:1}.  The morphology type data was obtained from the NED database\footnote{The NED database is available at: http://nedwww.ipac.caltech.edu. }.  The de Vaucouleurs radius $R_{25}$, the $m_b$ (``btc'' in the Leda database), the $m_i$ (``itc'' in the Leda database), and $W_{20}$ data came from the LEDA database \footnote{The LEDA database is available at: http://leda.univ-lyon1.fr. }.  The value of ``btc'' for close galaxies was used since I band data is generally unavailable.  In 35 of the 310 close galaxies, the ``btc'' was not available in LEDA.  The B band magnitude from NED, which is not corrected, was then used.  In 14 galaxies $m_b$ was unavailable from either database.  These galaxies were deleted from the calculation and the next closest galaxy added to have 10 galaxies for the calculation.  The $m_b$ for the galaxies which used the NED magnitude have relatively high values.  Therefore, their significance was low.  The $D_c$ was taken from \citet{free} distance modulus except as noted in Table \ref{tab:1}.

The sample of target galaxies included (A) low surface brightness (LSB), medium surface brightness (MSB), and high surface brightness (HSB) galaxies, (B) galaxies with a range of $W_{20}$ values from 61 $km s^{-1}$ to 607 $km s^{-1}$, (C) LINER, Sy, HII and less active galaxies, (D) galaxies which have excellent and poor agreement between $D_{tf}$ and $D_c$, (E) a distance range from 2.02 Mpc to 49.7 Mpc, (F) field and cluster galaxies, and (G) galaxies with rising, flat, and declining rotation curves.

For each of the target galaxies, the $D_{tf}$ was calculated using the equations from \citet{tull}.  The $W_{20}$ value was corrected for inclination, and used for $W^i_R$ in the TF equations.  For target galaxies, $D_{tf}$ was calculated using total apparent corrected I band magnitude (``itc'') and equations except for the galaxies IC 4182, NGC 1326A, NGC 4496A, and NGC 4571.  For these target galaxies, the ``btc'' and equations were used since ``itc'' data was unavailable in LEDA.

NGC 3031 and NGC 3319 are HSB galaxies with significant differences between $D_{tf}$ and $D_c$ measurements.  NGC 1365 has a rapidly declining rotation curve with a decline of at least 63\% of the peak value \citep{jors}.  NGC 2841 is a candidate to falsify the MOdified Newtonian Dynamics (MOND) model \citep{bott}.  A distance greater than 19 Mpc, compared to the measured $D_c$ of 14.1 Mpc, or a high mass/luminosity ratio is needed for MOND.  The $D_{tf}$ value of 25.5 Mpc given herein is compatible with MOND.  NGC 3031 has significant non-circular motion in its HI gas \citep{rots2}.  NGC 3198 is well suited for testing theories since the data for it has the largest radius and number of surface brightness scale lengths \citep{vana}.  NGC 4321 has a very asymmetric HI rotation curve and is ``lopsided''\citep{knap} in the disk region of the galaxy.

The NED database was used to assemble a list of galaxies closest to each of the target galaxies.  The selection of close galaxies used the Hubble Law with a Hubble constant $H_o$ of 70 $km s^{-1}Mpc^{-1}$ and galactic z to calculate the distances to galaxies.  The angular coordinates were taken from the NED database.  For each target galaxy, the distance from each target galaxy to each of the other galaxies was calculated.  The neighbor galaxies with the smallest distance were chosen as the close galaxies.

The uncertainty of the value of $H_o$ is high.  The use of $H_o$ herein is restricted to only the selection of the neighbor galaxies.  As the relative distances become larger, the distance uncertainty increases if only because of a greater variation in peculiar velocity.  Therefore, there is a practical upper limit on the number of galaxies that can be used to evaluate $E_a$.  If too few close galaxies are used, $E_a$ will be underestimated.  The number of close galaxies found to produce the highest correlation coefficient was 10.

For each close galaxy, the distance from the Milky Way to the galaxies $D_{g(jk)}$ (in $Mpc$) was calculated by,
\begin{equation}
D_{g( jk)} = \frac{z_{g( j)}}{z_{c_{( k)}}} D_{c( k)}
\label{eq:10},
\end{equation}
where $z_{g( j)}$ is the $z$ of the $j^{th}$ close galaxy, $z_{c( k)}$ is the $z$ of the $k^{th}$ target galaxy, and $D_{c( k)}$ is the $D_{c}$ of the $k^{th}$ target galaxy.  A bold character denotes a vector and the same character without the bold denotes the magnitude of the vector.  The sub $(  )$ indicated the letters enclosed are indices identifying the galaxies of the calculation.  The vector direction of $\bm D_{g( jk)}$ was calculated from coordinates found in the NED database.

The distance $R_{( jk)}$ from the $j^{th}$ close galaxy to the $k^{th}$ target galaxy  was calculated as, 
\begin{equation}
R_{( jk)} = \vert \bm R_{( jk)} \vert =  \vert \bm D_{c( k)} - \bm D_{g( jk)} \vert
\label{eq:11}.
\end{equation}

Calculating $D_{g(jk)}$ by the method of equation (\ref{eq:10}) reduces the error of $R_{( jk)}$ compared to other methods.

The total absolute magnitude $M_{(j)}$ of the $j^{th}$ close galaxy was calculated, 
\begin{equation}
M_{( j)} = m_b -25 - 5 \log (  D_{g( jk)} )
\label{eq:12}.
\end{equation}

\begin{figure}
\plotone{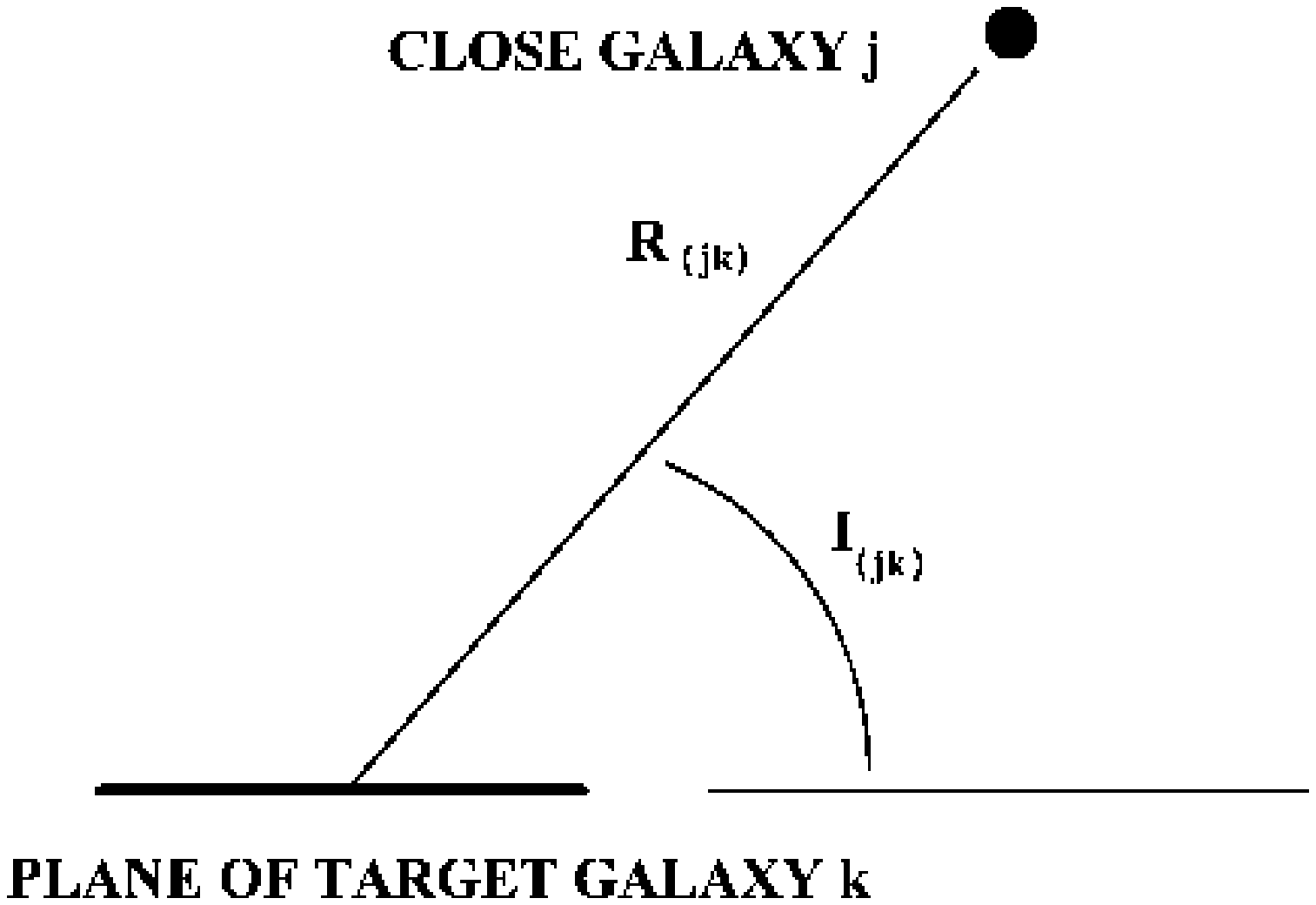}
\caption{\label{fig:3} Diagram showing relationship of $R_{(jk)}$ and $I_{(jk)}$.}
\end{figure}

\begin{figure}
\plotone{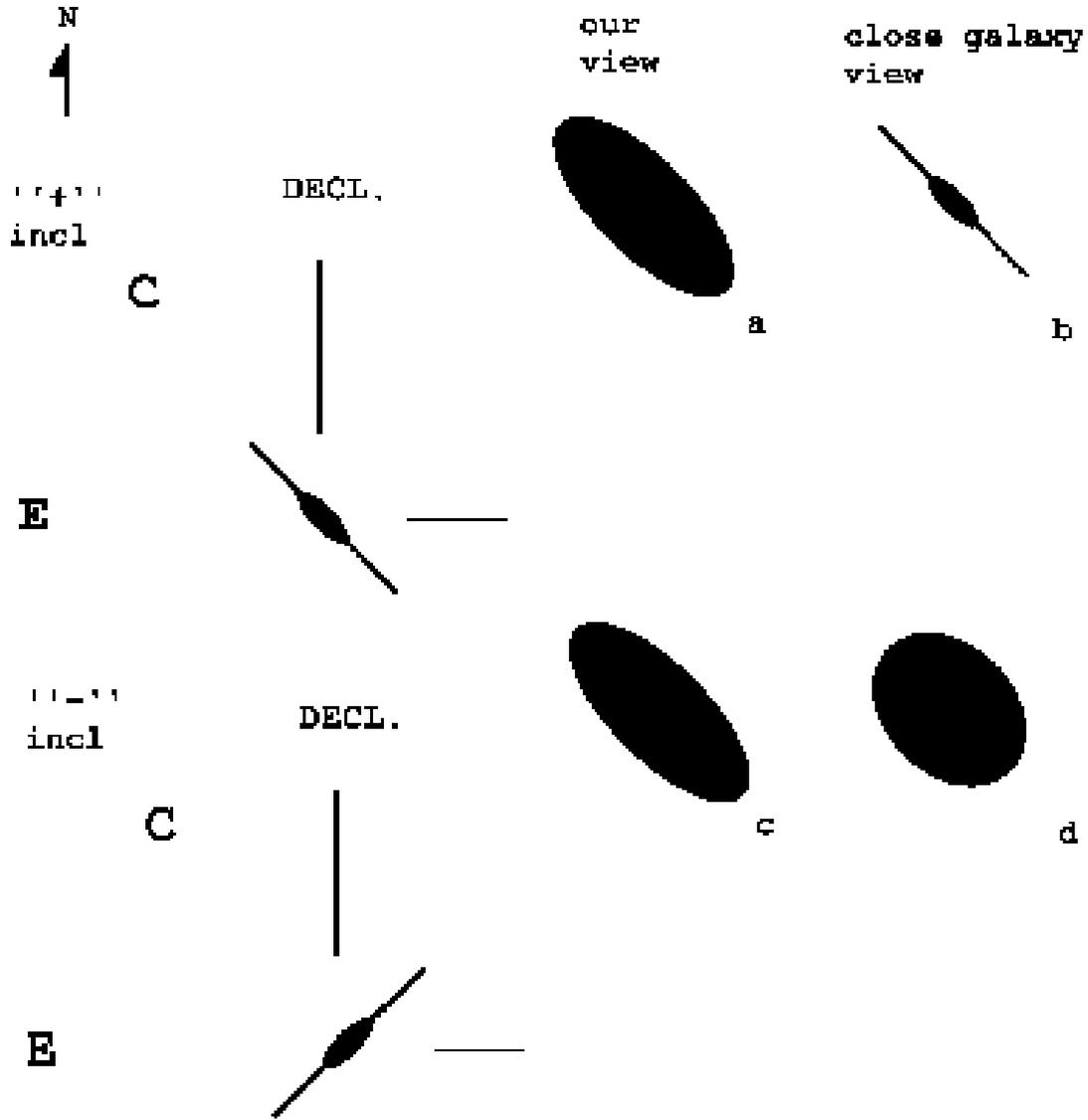}
\caption{\label{fig:4} Diagram showing relationship of the sign of inclination  (``incl'' in the figure) on our ( ``E'') view of the target galaxy compared to the $A_{(jk)}$ presented to a close (``C'') galaxy.  The axis labeled ``DECL.'' is the direction of the declination angle.  The other axis is the line of sight from earth.  The left diagrams show the 2 dimensional plane containing the 3 galaxies and the hypothetical relationship of the 3 galaxies.}
\end{figure}

The cross sectional area of the $k^{th}$ target galaxy $A$ presented to the $j^{th}$ close galaxy was calculated by,
\begin{equation}
A_{( jk)} = \pi R_{25}^2 \sin I_{( jk)}
\label{eq:14},
\end{equation}
where $I_{ ( jk ) }$ is the angle  between the $k^{th}$ target galaxy's plane and $\bm R_{( jk)}$ (see Figure~\ref{fig:3}) and is calculated from $i$ from the LEDA database (``incl''), the position angle ``pa'' from the LEDA database and the orientation sign as noted in Table \ref{tab:1} in the $i$ column.  The dependence of $A_{(jk)}$ on a target galaxy's orientation relative to a close galaxy may be significant.  The sign indicates whether the galaxy's polar axis is rotated clockwise or counterclockwise from our line of sight about the major axis when viewed from the easterly side of the major axis.  Figure~\ref{fig:4} is a diagram depicting a hypothetical example of our view and a close galaxy's view of the same galaxy.  Our view (a and c in Figure~\ref{fig:4}) is independent of whether the northwest side of the target galaxy is closer or farther than the southeast side.  The close galaxy will have one of two possible views depending on the orientation of the target galaxy.  In view b of Figure~\ref{fig:4} the close galaxy is located in the plane of the target galaxy with a positive $i$ and $A_{(jk)}$ is minimal.  With a negative $i$ sign, $A_{(jk)}$ as depicted in view d of Figure~\ref{fig:4} has a significant value.

The $E_a$ was calculated as,
\begin{equation}
E_{a( k)} = \sum_{j=1}^n 10^{\frac{- M_{( jk)}}{2.5}} \frac{A_{( jk)}}{4 \pi R_{( jk)}^2}  
\label{eq:15},
\end{equation}
where n is the number of galaxies included in the calculation.  In this case, $n = 10$.  The unit of measure of $E_a$ is in flux units ($erg \, cm^{-2} \, s^{-1}$).

The procedure to determine the sign of $i$ is as follows.
(A) The difference $\Delta E_a$ between the value of $E_a$ with a positive $i$ and $E_a$ with negative $i$ was calculated for each target galaxy.
(B) Nine target galaxies had a $\Delta E_a$ of less than 5 in the second significant figure, 17 had a $\Delta E_a$ of less than 1 in the first significant figure.  The sign of $i$ was of minimal significance for these galaxies.
(C) Since the relative position of these 17 galaxies on the $C_f$ versus $E_a$ plot changed little, these 17 target galaxies established the initial $E_a$ ranges and the initial constants of equation~(\ref{eq:9}).
(D) The sign of $i$ of the remaining target galaxies was determined such that their value of $E_a$ was within their error limits of one of the initial lines.
(E) The sign of $i$ of the initial 17 galaxies was re-chosen such that the value was closest to the line.
(F) The equations of the lines were recalculated.

The results are presented in Table~\ref{tab:1}.

\begin{figure}
\plotone{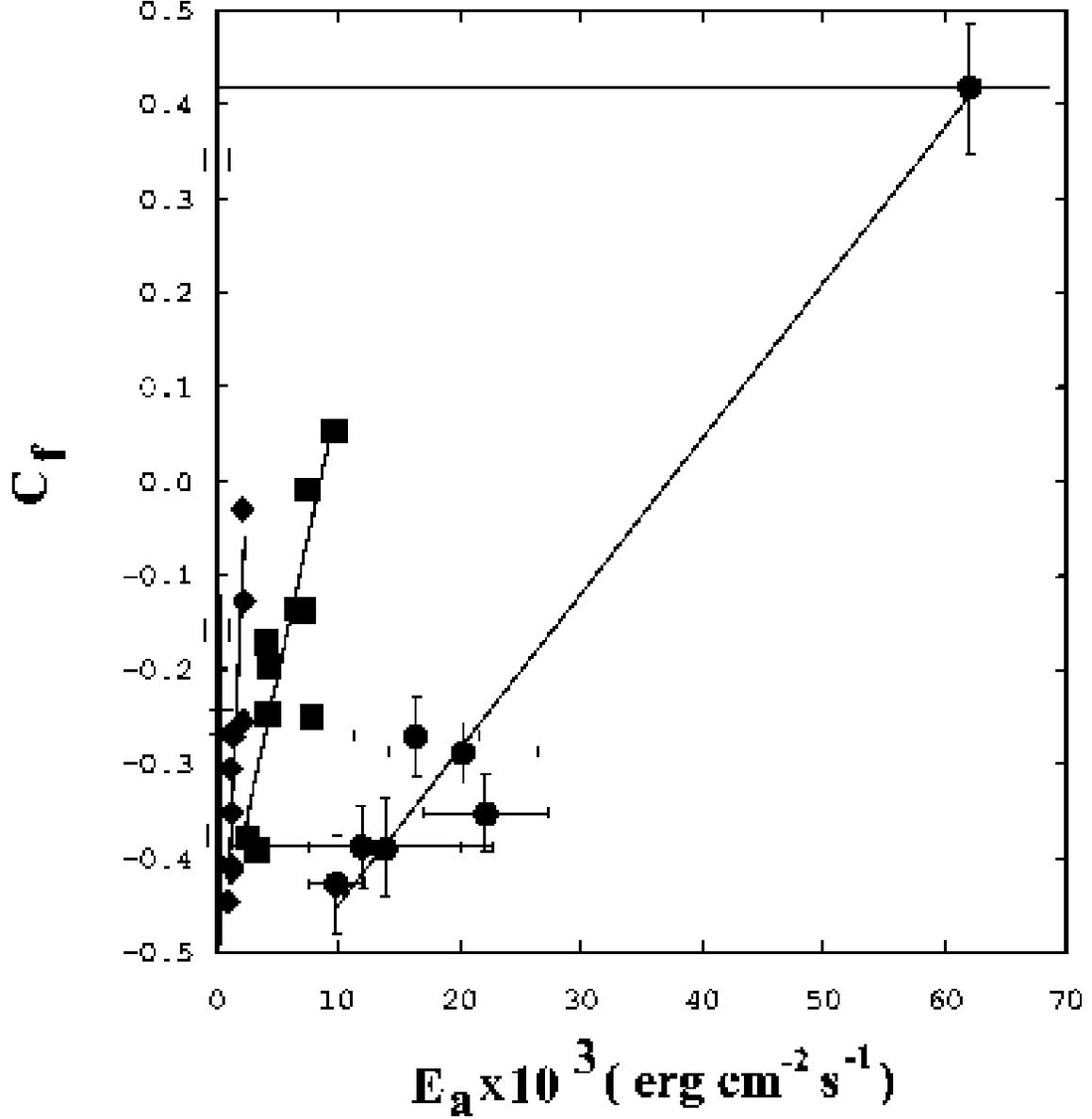}
\caption{\label{fig:1} Plot of $C_f$ versus $E_a$.  The relationship of the four zones is shown.  The lines are plots of the linear equation for each CET zone with the slopes and intercepts as listed in Table \ref{tab:2}.  The lines intersect at $( E_a , C_f ) = ( 550 \pm 50 erg \, cm^{-2} \, s^{-1}, -0.54 \pm 0.07 )$.  Open squares designates the galaxies in CET 1, filled diamonds designate galaxies in CET 2, filled squares designate galaxies in CET 3, and filled circles designate galaxies in CET 4.}
\end{figure}

\begin{figure}
\plotone{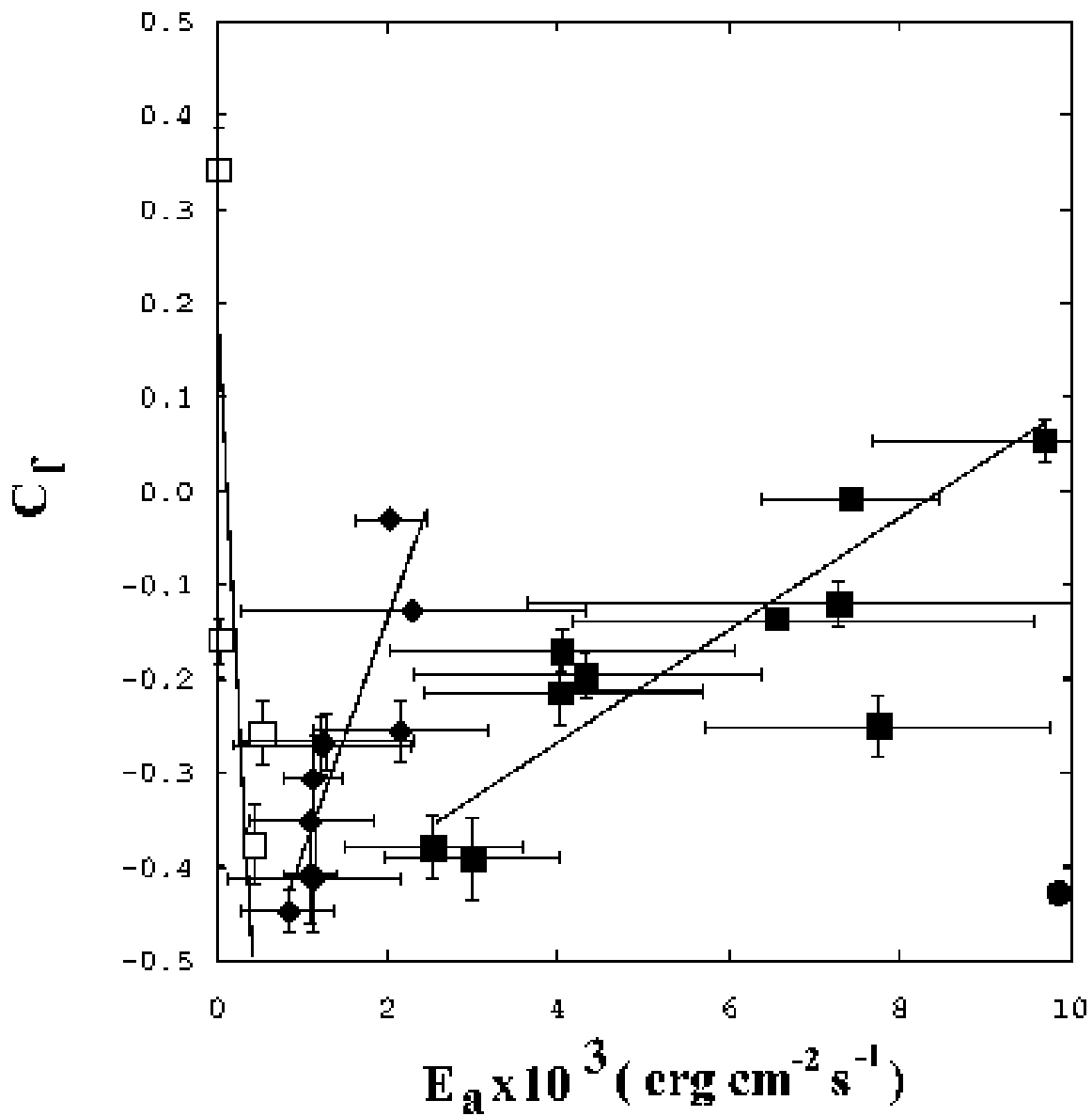}
\caption{\label{fig:2} Plot of $C_f$ versus $E_a$.  This is an expanded view of the first 3 CET zones.  The relationship of the four zones is shown in Figure~\ref{fig:1}.  The lines are plots of the linear equation for each CET zone with the slopes and intercepts as listed in Table \ref{tab:2}.  Open squares designates the galaxies in CET 1, filled diamonds designate galaxies in CET 2, filled squares designate galaxies in CET 3, and filled circles designate galaxies in CET 4. }
\end{figure}

A plot of $C_f$ versus $E_a$ is shown in Figures~\ref{fig:1} and \ref{fig:2}.  The galaxies form into four distinct Conversion Efficiency Type (CET) relations, labeled CET 1, CET 2, CET 3, and CET 4.  The characteristics of each CET is tabulated in Table~\ref{tab:2}.  The statistical test of variance of the line and $E_a$ of the sample galaxies (F test ) for all CET regions is 0.99.  The projections of the lines intersect at ($E_a$, $C_f$) = (550$\pm$50 $ erg \, cm^{-2} \, s^{-1}$,-0.54$\pm$.07).

The error $\partial E_a ( D_c )$ in $E_a$ due solely to the error in $D_c$ is also listed in Table~\ref{tab:1} for each target galaxy.  Table~\ref{tab:2} shows the range of $E_a$ and $\partial E_a ( D_c )$ for each CET.  The choice of sign of $i$ could also have been done using the criteria that values of $E_a$ and $\partial E_a ( D_c )$ be within the ranges specified in Table~\ref{tab:2}.

The large error in NGC 4603 is a result of the unusually large published error in $D_c$.  NGC 4535 was the only galaxy where the line was outside its error limits.  NGC 4535 could fit nearly as well in CET 4 with a positive $i$ sign and an out of range $\partial E_a ( D_c )$.  If the sign of $i$ of NGC 4535 is positive, NGC 4535 is closer to CET 4, the correlation coefficient of CET 3 increases to 0.94, and the $\partial E_a ( D_c )$ is in the CET 3 range.  Therefore, if CET 4 is selected for NGC 4535, the $\partial E_a ( D_c )$ and the $E_a$ are in different ranges.  Hence, NGC 4535 was considered to be in CET 3.

The analysis presented herein depends on the Cepheid distances being secure, standard distances.  \citet{free} adopted a metallicity PL correction factor of $-0.2 \pm 0.2 \,mag \, dex^{-1}$.  \citet{ferr} assumed the PL relation was independent of metallicity.  If the Cepheid distances used are from the earlier work of \citet{ferr}, which are higher than \citet{free}, the correlation coefficients are ~-0.74, 0.89, 0.85, and 0.98 for the CET 1, CET 2, CET 3, and CET 4 zones, respectively, and the slopes and intercepts of the new equations are within 3 $\sigma$ of the slopes and intercepts of the equations of Table~\ref{tab:2}.  Thus, the theory and the result presented herein appear robust relative to systematic variation in the Cepheid distances.

The analysis also depends on the Hubble law.  There is considerable uncertainty in the value of $H_o$.  The attempt to minimize this uncertainty was to consider only galaxies near the target galaxy and to calculate the distances between galaxies using the ratio of $z$ values.  This assumes that the $H_o$ varies little around each target galaxy.  The effect of the error in $H_o$ is limited to a selection bias.  Using $H_o = 80 km s^{-1}Mpc^{-1}$ produced no significant change in the results.

NGC 3621, NGC 3319, and NGC 4535 are outliers in CET 1, CET 2, and CET 3, respectively.  These three galaxies have a more negative $C_f$ than their respective theoretical lines.  The removal of these galaxies from consideration yields the modified correlation coefficient, modified $K_s$, and modified $K_i$ listed in Table~\ref{tab:3}.

\section{DISCUSSION}

\citet{shan2} suggested the Cepheid distances may be underestimated due to metallicity and magnitude incompleteness.  Also, the sign and value of a possible metallicity correction factor is uncertain \citep{free}.  As the difference between \citet{ferr} and \citet{free}, the effect will be to change only the slope and intercept of the CET equations.

The existence of four $E_a$ zones is unexplained.  It should be noted that for each zone $K_s$ and $K_i$ are constants distinct from other zones.  We speculate that since $K_s$ has discrete values, equation (\ref{eq:9a}) implies $K_a$, $K_e \epsilon$,  or both must have discrete values, also.  That is, the values are not continuously variable.  If $K_a$ and $K_e \epsilon$ depend on internal characteristics of a galaxy such as chemical composition, then the gap between CET zones and a discrete value $K_s$ implies $E_a$ must influence at least two of $K_a$, $K_e$, or $\epsilon$.  We speculate that other galaxies influence parameters internal to a galaxy including chemical composition.  Another unexplained characteristic is the ratio of the slopes.  The ratio of the slopes is 4.24 between CET 3 and CET 2 and 3.65 between CET 4 and CET 3 (both approximately 4).  Also unexplained is the convergence of the linear equations to ($E_a$, $C_f$) = (550$\pm$50 $ erg \, cm^{-2} \, s^{-1}$,-0.54$\pm$.07).

Since the $C_f - E_a$ relationship has excellent correlation coefficients without NGC 3621, NGC 3319, and NGC 4535, some of the assumptions made in developing the model may be invalid for these galaxies.  NGC 3319 is rich in HI \citep{moor} and NGC 4535 is poor in HI \citep{giov}.  However, if a galaxy has different absorption due to HI, then the $K_e \epsilon$ should also change since HI is a fuel for emission, also.  Another intriguing possibility is that there are three more CET zones.  The additional zones would consist of (A) NGC 2090 and NGC 3031, (B) NGC 1326A and NGC 3319, and (C) NGC 1365 and NGC 4535.  The functional relationship would be maintained.  Another method to determine the orientation (sign) of each galaxy, more data, and reduced error are needed.

The choice of sign of the inclination angle is the most subjective part of the data.  In some cases such as NGC 0300, the difference in $E_a$ causes a change of CET zone which leaves the $C_f$ nearly the same.  As expected, a sign change has greater effect on the galaxies with one very close neighbor.  If all galaxies have their inclination sign reversed, four CET categories are again formed with correlation coefficients of -0.72, 0.98, 0.91, and 0.93.  However, several galaxies are outside their error limits of a line.  The method is robust.  The choice of sign reported herein allows this method to work.  However, a better method in determining the sign would place greater confidence in the numbers.

NGC 2841 is close to the calculated line in CET 2 with a measured $C_f =0.45$ and a calculated $C_f =0.41$.  Therefore, the model in this Paper accounts for nearly all the error in ($D_c - D_{tf}$) and $D_c$ is much less than MOND requires.  This suggests NGC 2841 is a problem for MOND.

When a galaxy's fundamental plane parameter is compared to the galaxy's luminosity, a broader scatter and outliers (see, for example, \citet{gebh}) result than if a parameter other than luminosity is compared.  This paper suggests this is due to the effect of neighboring galaxies.

Relative to the TF equations, the model must still discover the relationship of $W_{20}$ to the $C_f$.  We speculate that after the adjustment for neighboring galaxies, this correction may explain the scale error found by \citet{shan2} between raw Cepheid and TF distances.  The TF assumption that the intrinsic mass to intrinsic luminosity ratio is constant among galaxies \citep{aaro} is valid for $z < 0.006$.  The TF approach can be an even more powerful tool with $C_f$ considered.

\section{CONCLUSION}

$\bullet$~The sample considered 31 spiral galaxies with published Cepheid distances.  Predicted linear relationships between a factor expressing the relation between TF and Cepheid distance calculations and a factor expressing the amount of impinging energy from nearby galaxies were found.

$\bullet$~The impinging energy relation was found to have four zones depending on the amount of impinging energy.  Each zone had a different linear relationship to the distance correction factor.

$\bullet$~The lines from the four zones projects to a common point at ($E_a$, $C_f$) = (550$\pm$50$erg \, cm^{-2} \, s^{-1}$,-0.54$\pm$.07).

$\bullet$~When considering the luminosity of a galaxy, the effect of its neighbors should be considered.

$\bullet$~The intrinsic mass to intrinsic luminosity ratio is constant among the sample galaxies. 

$\bullet$~The mass is constant for each of the target galaxies.  Therefore, the intrinsic luminosity equals the rate of energy conversion of the mass being added to the galaxy.

\acknowledgments
This research has made use of the NASA/IPAC Extragalactic Database (NED), which is operated by the Jet Propulsion Laboratory, California Institute of Technology, under contract with the National Aeronautics and Space Administration.

This research has made use of the LEDA database (http://leda.univ-lyon1.fr).

We thank W. Osborne and K. Rumstay for comments and suggestions that improved the manuscript.

We acknowledge and appreciate the financial support of Cameron Hodge, Stanley, New York, and Maynard Clark, Apollo Beach, Florida, while working on this project.

\clearpage

\clearpage

\clearpage

\clearpage

\clearpage

\begin{deluxetable}{llrrrrrrr}
\tabletypesize{\scriptsize}
\tablecaption{\label{tab:1} Data for the target galaxies.}
\tablehead{\colhead{Galaxy} &\colhead{NED type \tablenotemark{a}} &\colhead{i \tablenotemark{b}} &\colhead{$m_i$ \tablenotemark{c}} &\colhead{$M_i$ \tablenotemark{d}}&\colhead{$D_{tf}$ \tablenotemark{e}} &\colhead{$D_c$\tablenotemark{f}} &\colhead{$E_a$\tablenotemark{g}} &\colhead{$\partial E_a ( D_c )$ \tablenotemark{h}}} 
\startdata
CET 1&\\
\objectname[]{NGC 2090} &SA(rs)b HII&$-$68.3&9.35&-21.57&15.26$\pm$2.26&11.44$\pm$\phm{6}0.21&500$\pm$\phm{000}500&20\\
\objectname[]{NGC 3031} &SA(s)ab;LINER Sy1.8&$-$59.0&5.41&-23.28&5.48$\pm$0.78&3.55$\pm$\phm{6}0.13&400$\pm$\phm{000}100&30\\
\objectname[]{NGC 3621} &SA(s)d&$-$65.6&8.01&-21.43&7.72$\pm$1.15&6.55$\pm$\phm{6}0.18&53$\pm$\phm{0000}40&4\\
\objectname[]{NGC 5253} &Im pec;HII Sbrst&$+$66.7&9.40&-17.50&2.40$\pm$0.36&3.25$\pm$\phm{6}0.22&13$\pm$\phm{00000}5&1\\
\\
CET 2\\
\objectname[]{IC  4182} &SA(s)m&$-$27.1&12.27\tablenotemark{i}&-17.13\tablenotemark{i}&7.60$\pm$1.13&4.53$\pm$\phm{6}0.13&1100$\pm$\phm{000}300&50\\
\objectname[]{NGC 0300} &SA(s)d&$-$39.8 &7.23&-20.45&3.44$\pm$0.50&2.02$\pm$\phm{6}0.07&1000$\pm$\phm{00}1000&60\\
\objectname[]{NGC 0925} &SAB(s)d;HII&$-$61.1&9.19&-20.68&9.41$\pm$1.39&9.13$\pm$\phm{6}0.17&2000$\pm$\phm{000}400&90\\
\objectname[]{NGC 1326A} &SB(s)m&$-$42.5&14.26\tablenotemark{i}&-17.04\tablenotemark{i}&18.20$\pm$2.70&16.15$\pm$\phm{6}0.77&2400$\pm$\phm{00}2000&200\\
\objectname[]{NGC 2403} &SAB(s)cd&$-$60.0&7.32&-21.10&4.84$\pm$0.71&3.14$\pm$ 0.14&1100$\pm$\phm{000}800&200\\
\objectname[]{NGC 2541} &SA(s)cd LINER&$+$66.8&10.53&-20.40&15.38$\pm$2.28&11.23$\pm$\phm{6}0.26&1200$\pm$\phm{00}1000&80\\
\objectname[]{NGC 2841} &SA( r)b;LINER Sy&$+$68.0&7.91&-24.12&25.51$\pm$3.78&14.07$\pm$\phm{6}1.57\tablenotemark{j}&900$\pm$\phm{000}600&200\\
\objectname[]{NGC 3319} &SB(rs)cd;HII&$-$59.1&10.39&-20.89&18.05$\pm$2.78&13.44$\pm$\phm{6}0.57&2100$\pm$\phm{00}1000&200\\
\objectname[]{NGC 4414} &SA(s)c  LINER&$-$54.0&8.87&-23.23&26.33$\pm$3.90&16.61$\pm$\phm{6}0.38 &900$\pm$\phm{00}1000&100\\
\objectname[]{NGC 4548} &SBb(rs);SY LINER&$+$37.0&8.89&-22.78&21.60$\pm$3.20&15.01$\pm$\phm{6}0.35&1200$\pm$\phm{000}300&50\\
\\
CET 3\\
\objectname[]{NGC 1365} &SBb(s)b Sy1.8&$+$57.7&8.18&-22.91&16.51$\pm$7.55&17.23$\pm$\phm{6}0.40&9500$\pm$\phm{00}2000&400\\
\objectname[]{NGC 1425} &SA(rs)b&$+$69.5&9.56&-22.30&23.57$\pm$3.49&20.91$\pm$\phm{6}0.48&7500$\pm$\phm{00}3000&300\\
\objectname[]{NGC 3198} &SB(rs)c&$-$70.0&9.22&-21.82&16.19$\pm$2.40&13.69$\pm$\phm{6}0.51&4000$\pm$\phm{00}2000&300\\
\objectname[]{NGC 3351} &SB(r)b HII Sbrst&$-$41.5&8.32&-22.58&15.15$\pm$2.25&9.34$\pm$\phm{6}0.39&3100$\pm$\phm{00}1000&400\\
\objectname[]{NGC 3627} &SAB(s)b;Sy2  LINER&$+$57.3&7.54&-22.79&11.62$\pm$1.73&9.38$\pm$\phm{6}0.35&4900$\pm$\phm{00}2000&400\\
\objectname[]{NGC 4258} &SAB(s)bc;Sy1.9  LINER&$-$72.0&7.04&-22.87&9.62$\pm$1.42&7.73$\pm$\phm{6}0.26&4400$\pm$\phm{00}2000&300\\
\objectname[]{NGC 4496A} &SB(rs)m&$+$48.1&11.78\tablenotemark{j}&-19.04\tablenotemark{i}&14.58$\pm$2.16&14.53$\pm$\phm{6}0.20&7500$\pm$\phm{00}1000&300\\
\objectname[]{NGC 4535} &SAB(s)c&$-$44.0&8.89&-22.57&19.64$\pm$2.91&14.80$\pm$ 0.35&7900$\pm$\phm{00}2000&300\\
\objectname[]{NGC 4571} &SA(r)d&$-$30.0&11.6\tablenotemark{j}&-20.31\tablenotemark{i}&24.06$\pm$3.57&15.15$\pm$\phm{6}1.46\tablenotemark{k} &2600$\pm$\phm{00}1000&400\\
\objectname[]{NGC 7331} &SA(s)b;LINER&$-$75.0&7.70&-23.43&16.82$\pm$2.49&14.53$\pm$\phm{6}0.62&6700$\pm$\phm{00}3000&600\\
\\
CET 4&\\
\objectname[]{NGC 3368} &SAB(rs)ab;Sy LINER&$+$54.7&7.99&-22.70&13.76$\pm$2.04&9.87$\pm$\phm{6}0.28&21000$\pm$\phm{00}7000&2000\\
\objectname[]{NGC 4321} &SAB(s)bc; LINER HII&$-$30.0&8.39&-23.44&23.23$\pm$3.44&14.33$\pm$\phm{6}0.47&14000$\pm$\phm{00}7000&900\\
\objectname[]{NGC 4536} &SAB(rs)bc&$-$58.9&9.15&-22.32&19.66$\pm$2.91&14.46$\pm$\phm{6}0.27&17000$\pm$\phm{00}5000&700\\
\objectname[]{NGC 4603} &SA(rs)bc&$-$55.0&9.86&-22.86&35.05$\pm$5.20&49.70$\pm$24.15\tablenotemark{l}&62000$\pm$400000&80000\\
\objectname[]{NGC 4639} &SAB(rs)bc;Sy1.8&$-$52.0&10.27&-22.54&36.45$\pm$5.39&21.00$\pm$\phm{6}0.79&10000$\pm$\phm{00}2000&600\\
\objectname[]{NGC 4725} &SAB(r)ab;Sy2 pec&$-$54.4&8.04&-23.24&18.03$\pm$2.67&11.92$\pm$\phm{6}0.33&24000$\pm$\phm{00}5000&1000\\
\objectname[]{NGC 5457} &SAB(rs)cd&$+$22.0&6.99&-23.18&10.82$\pm$1.60&6.70$\pm$\phm{6}0.35 &12000$\pm$\phm{0}11000&1000\\
\enddata

\tablenotetext{a}{Galaxy morphological type from the NED database.}
\tablenotetext{b}{Galaxy inclination between line of sight and polar axis (in degrees) from the LEDA database.  The $+$ or $-$ sign signifies galaxy inclination orientation.}
\tablenotetext{c}{ Total apparent corrected I-magnitude ``itc'' of the $i^{th}$ galaxy from the LEDA database, unless otherwise noted.}
\tablenotetext{d}{The absolute I-magnitude $M^{b,k,i}_I$ from \citet{tull}, unless otherwise noted.}
\tablenotetext{e}{The calculated distance from (m-M) in Mpc.}
\tablenotetext{f}{Distance in Mpc from Cepheid data from \citet{free}, unless otherwise noted.}
\tablenotetext{g}{In flux units $erg \, cm^{-2} \, s^{-1}$.}
\tablenotetext{h}{The $E_a$ error due solely to $D_c$ error.}
\tablenotetext{i}{Magnitude data is total apparent corrected B-magnitude ``btc'' from the LEDA database and $M^{b,k,i}_B$ from \citet{tull}.}
\tablenotetext{j}{Distance is from \citet{macr}.}
\tablenotetext{k}{Distance is from \citet{pier}.}
\tablenotetext{l}{Distance is from \citet{patu}.}

\end{deluxetable}

\clearpage

\begin{deluxetable}{lcrrrrr}
\tabletypesize{\scriptsize}
\tablecaption{\label{tab:2} Data for the Conversion Efficiency Types.}
\tablehead{\colhead{} &\colhead{No. \tablenotemark{a} } &\colhead{$E_a$\tablenotemark{b} } &\colhead{$\partial E_a ( D_c )$\tablenotemark{c} } &\colhead{Corr. \tablenotemark{d} } &\colhead{$K_s$ \tablenotemark{e} } &\colhead{$K_i$ \tablenotemark{f} }
}
\startdata
CET 1&4&0 - \phm{0}800&0 - \phm{00}40&-0.77&$-1.3 ( \pm 0.6 ) \times 10^{-3}$&0.2\phm{0}($\pm$0.2\phm{0})\\
CET 2&10&801 - 2400&40 - \phm{0}300&0.87&$2.5 ( \pm 0.5 ) \times 10^{-4}$&-0.63($\pm$0.07) \\
CET 3&10&2401 - 9700&300 - \phm{0}600&0.81&$6.0 ( \pm 1.0 ) \times 10^{-5}$&-0.51($\pm$0.08) \\
CET 4&7&9700 - \phm{0000}&600 - \phm{0000}&0.98&$1.6 ( \pm 0.1 ) \times 10^{-5}$&-0.61($\pm$0.04)\\

\enddata

\tablenotetext{a}{The number of galaxies of the sample in each CET.}
\tablenotetext{b}{The $E_a$ range of the CET in flux units of $erg \, cm^{-2} \, s^{-1}$.}
\tablenotetext{c}{The $E_a$ error range due solely to $D_c$ error of the CET in flux units of $erg \, cm^{-2} \, s^{-1}$.}
\tablenotetext{d}{Correlation Coefficient.}
\tablenotetext{e}{The best fit slope in units of $erg^{-1} \, cm^{2} \, s^{1}$ derived to make the F test greater than 0.99.  The error is 1 $\sigma$. }
\tablenotetext{f}{The best fit intercept derived to minimize the total difference between data and the line.  The error is 1 $\sigma$.}

\end{deluxetable}

\clearpage

\begin{deluxetable}{lrrr}
\tabletypesize{\scriptsize}
\tablecaption{\label{tab:3} Data for the Conversion Efficiency Types without NGC 3621, NGC 3319, and NGC 4535.}
\tablehead{\colhead{} &\colhead{MOD Corr. \tablenotemark{a} } &\colhead{MOD $K_s$ \tablenotemark{b} } &\colhead{MOD $K_i$ \tablenotemark{c} }
}
\startdata
CET 1&-0.98&$-1.5 ( \pm 0.3 ) \times 10^{-3}$& 0.38($\pm$0.10)\\
CET 2&0.92& $ 2.7 ( \pm 0.4 ) \times 10^{-4}$&-0.66($\pm$0.06)\\
CET 3&0.93& $ 6.2 ( \pm 0.9 ) \times 10^{-5}$&-0.52($\pm$0.05)\\
CET 4&0.98& $ 1.6 ( \pm 0.1 ) \times 10^{-5}$&-0.61($\pm$0.04)\\

\enddata

\tablenotetext{a}{Correlation Coefficient.}
\tablenotetext{b}{The best fit slope in units of $erg^{-1} \, cm^{2} \, s^{1}$ derived to make the F test greater than 0.99.  The error is 1 $\sigma$. }
\tablenotetext{c}{The best fit intercept derived to minimize the total difference between data and the line.  The error is 1 $\sigma$. }

\end{deluxetable}

\clearpage

\end{document}